\documentstyle[preprint,aps]{revtex}
\begin{document}
\draft
\preprint{RU9544}
\title{Atom in a $q$-Analog Harmonic Oscillator Trap}
\author{S. Shelly Sharma\thanks{email:sshelly@positron.rutgers.edu}}
\address{Department of Physics and Astronomy, Rutgers University,
Piscataway, NJ 08855-0849 USA, and \\
Departamento de Fisica, Universidade Estadual de Londrina\\
Londrina, Parana, 86051-970, Brazil}
\author{N. K. Sharma}
\address{Department of Mathematics, Rutgers University\\ New Brunswick,
NJ, USA\\
and\\
Departamento de Matematica, Universidade Estadual de Londrina\\
Londrina, Parana, 86051-970, Brazil}
\author{Larry Zamick}
\address{Department of Physics and Astronomy, Rutgers University, \\
Piscataway NJ 08855-0849, USA}

\date{\today}

\maketitle
\begin{abstract}
We study the population inversion and Q-function
of a two-level atom, interacting with
single-mode laser light field, in a $q$-analog harmonic oscillator
trap for increasing $q$.
For $\tau=.003(q=e^{\tau})$ the
collapses and revivals of population inversion
become well defined facilitating
experimental
observation but for large
$\tau \sim 0.1$ the time dependence
of population inversion is completely wiped out.

\end{abstract}

\Roman{section}
\newpage

\narrowtext

\section{Introduction}

\hspace{0.5 in}
Recent development of quantum groups \cite{Jim,Woron,Pasq} has resulted in
the construction of $q$-analogs of quite a few quantum systems including a very
popular system that is the harmonic oscillator. A $q$-analog realization
of harmonic oscillator has been given by Macfarlane\cite{Mac89} and
Biedenharn\cite{Bied89}. In Ref. \cite{Bied89} coherent analogs of Glauber
states have also been proposed. It is of great interest to use these quantum
analog formulations to study physical systems and there are situations where
a description in terms of $q$- analog structures is apparently more natural.
The case of an atom being cooled through interaction with the radiation
offers an interesting case study. It has been shown \cite{Died89} recently
that the atom can be cooled down to energies very close to it's zero point
vibrational energy. At these energies the quantum effects due to center of
mass motion of the atom are expected to be very important\cite{Javan81}.
Blockley et. al.\cite{block} have studied the collapses and revivals of
population inversion in a single two-level atom interacting with a classical
single mode travelling light field while constrained to move in a
one-dimensional harmonic Oscillator trap. They also discuss the possibility
of observing the collapses and revivals experimentally.

\hspace{0.5 in}
The interaction of the atom with the trap potential results in
energy exchange between the internal degrees of freedom of the atom
and the center of mass motion.
As the
atom slows down, the amount of energy exchanged in each step of the
cooling
process is not expected to be constant but a
variable dependent on the initial energy state of the atomic
center of mass. Presently we study the
time evolution of a
two-level
atom in a $q$-deformed harmonic oscillator trap, in interaction
with a single mode travelling light field. In a $q$-deformed harmonic
oscillator trap, the energy spacing between the trap
states, as seen by the atom,  is a function of
the initial vibrational state of the atom.
As such, when there is a lot of energy associated with center
of mass motion energy loss and gain occurs in large energy quanta.
However as the atom cools down, the energy exchange takes place
in smaller units of energy.
Besides that the relative energy separation between successive states is
determined by the value of the deformation parameter.
Blockley et al\cite{block} have
shown that in the Lamb-Dicke regime, when only the interaction between
nearest neighbours is significant the model is similar to the
Jaynes-Cummings Model(JCM)\cite{Jaynes} with trap quanta playing a role
similar to that of the light quanta in JCM. An atom with its center of mass
in a coherent trap state initially, shows collapses and revivals of its
population inversion. Our object is to investigate the system
response for a trap potential expected to be closer to experimental
situation in comparoson with a harmonic oscillator trap.
The energy spacings between energy eigen states of the $q$-analog harmonic
oscillator are not constant but are a function of $q$-deformation. As such for
a given system a suitable choice of the deformation parameter can take us
from the classical limit where the trap states are closely spaced to the
Lamb- Dicke regime with well spaced trap states. In our earlier attempts
\cite{She92,She94} to understand
the physical nature of deformation in the context of
pairing of nucleons, it has been found that the deformation
amounts to simulating the nonlinearities of the problem or part of the
interaction not included in the hamiltionian. With these results in mind we
expect the  -atom in $q$ deformed oscillator trap model-
to be a better description of the
physical situation involved.

\section{{\lowercase{$q$}}-Analog Harmonic Oscillator Trap}

\hspace{.5 in} Consider a single two level atom having atomic transition
frequency$\ \omega _a$ in a quantized $q$-analog quantum harmonic oscillator
trap($q$-deformed harmonic oscillator trap) interacting with a single
mode
travelling light field. The creation and annihilation
operators for the trap quanta satisfy the following quocommutation relations,

\begin{equation}
\label{comm}aa^{+}-qa^{+}a=q^{-N}\medskip\ ;\medskip\ \ \ Na^{+}-a^{+}N=a^{+}%
\medskip\ ;\medskip\ Na-aN=-a
\end{equation}

Here N is the number operator. The operators $a$ and $a^{+}$ act in a
Hilbert space with basis vectors $\left| n\right\rangle $, $n=0,1,2,...,$
given by,

\begin{equation}
\label{vec}\left| n\right\rangle =\frac{(a^{+})^n}{([n]_q!)^{\frac 12}}%
\left| 0\right\rangle
\end{equation}
such that $N\left| n\right\rangle =n\left| n\right\rangle .$ The vacuum
state is $a\left| 0\right\rangle =0.$
We define here $[x]_q$ as
\begin{equation}
\label{defx}[x]_q=\frac{q^x-q^{-x}}{q-q^{-1}}
\end{equation}
and the $q$-analog factorial $[n]_q!$ is recursively defined by

 [0]$%
_q!=[1]_q!=1$ and
$[n]_q!=[n]_q[n-1]_q!.$ It is easily verified that
\begin{equation}
\label{aaop}a^{+}\left| n\right\rangle =[n+1]_q^{\frac 12}\left|
n+1\right\rangle \medskip\ ;\medskip\ a\left| n\right\rangle =[n]_q^{\frac
12}\left| n-1\right\rangle
\end{equation}
and $N$ is not equal to $a^{+}a.$ Analogous to the harmonic oscillator one may
define the $q$-momentum and $q$-position coordinate
\begin{equation}
\label{pq}P_q=i\sqrt{\frac{m\hbar \omega }2}(a^{+}-a)\medskip\ ;\medskip\
Q_q=\sqrt{\frac \hbar {2m\omega }}(a^{+}+a).
\end{equation}
The $q$-analog harmonic oscillator hamiltonian is given by
\begin{equation}
\label{ham}H_{qho}=\frac 12\hbar \omega (aa^{+}+a^{+}a)
\end{equation}
with eigenvalues
\begin{equation}
\label{eig}E_{n}=\frac 12\hbar \omega ([n+1]_q+[n]_q).
\end{equation}

For $\tau\ll 1$ we have for the energy spacing
\begin{equation}
\label{spacing}E_{n+1}-E_{n}
\sim\hbar \omega +(n+1)^2 \tau^2 \hbar \omega.
\end{equation}

As the atomic state changes by absorption or emission of a light quantum
there is a change in it's vibrational state in the trap as well. We note
that the trap states are not evenly spaced, the energy spacing being a
function of deformation. Besides that as we move up in the number of
vibrational quanta in the states the spacing between successive states
increases. For a given region of vibrational states a suitable
choice of
deformation parameter should correspond to
the Lamb Dicke regime.
\section{DYNAMICS}

\hspace{.5 in}
The Hamiltonian for the system that consists the atom vibrating in the
trap and interacting
with a classical single-mode light field of frequency $\omega _l.$ is given
by\cite{block}
\begin{equation}
\label{Atomh}H=\frac 12\hbar \omega (a^{+}a+aa^{+})+\frac 12\hbar \triangle
\sigma _z+\frac 12\hbar \Omega (F\sigma ^{+}+F^{*}\sigma ^{-})
\end{equation}
where $\Delta =\omega _a-\omega _l$ , is the detuning parameter and $\Omega $
is the Rabi frequency of the system. The operator $F$ stands for $\exp
(ik Q_q)=\exp [i\epsilon (a^{+}+a)].$ The parameter $\epsilon =\sqrt{\frac{%
E_r}{E_t}}$ is a function of the ratio of the recoil energy of the atom $%
E_r= \frac{\hbar ^2k^2}{2m}$and the characteristic trap quantum energy $%
E_t=\hbar \omega $ in the limit $q\rightarrow 1.$ Here $k$ is the wave
vector of the light field. The second term in the Hamiltonian refers
to the energy associated with internal degrees of freedom of the atom,
whereas the third term is the interaction of the atom with the light field.
For nonzero values of
deformation parameter $q$, the kinetic energy and potential energy
of the atom in the trap are
a function of deformation. In addition the interaction of
the atom with the light field is also deformation dependent.

\hspace{.5 in}
The state of the system at a time t,
\begin{equation}
\label{psi}\Psi (t)=\sum\limits_mg_m(t)\left| g,m\right\rangle
+\sum\limits_me_m(t)\left| e,m\right\rangle
\end{equation}
is a solution of the time dependent Schrodinger equation
\begin{equation}
\label{seq}H\Psi (t)=i\hbar \frac d{dt}\Psi (t).
\end{equation}
The vector,
$\vert g,m \rangle$, corresponds to the atom being in its ground state
with its center of mass in m-th trap state. The label e stands for the
excited
state of the atom.

\hspace{.5 in}
The probability amplitudes $g_m$and $e_m$ satisfy the following set of
coupled equations

\begin{equation}
\label{gmdot}i\frac{dg_m}{dt}=\frac 12g_m(t)\left( \omega
([m+1]_q+[m]_q)-\Delta \right) +\frac 12\Omega
\sum\limits_ne_n(t)\left\langle g,m\right| \sigma ^{-}F^{*}\left|
e,n\right\rangle
\end{equation}

\begin{equation}
\label{emdot}\frac {de_m}{dt}=\frac 12e_m(t)\left( \omega
([m+1]_q+[m]_q)+\Delta \right) +\frac 12\Omega
\sum\limits_ng_n(t)\left\langle e,m\right| F\sigma ^{+}\left|
g,n\right\rangle
\end{equation}

\hspace{.5 in}
In order to evaluate the matrix elements $\left\langle m\right| F^{*}\left|
n\right\rangle =\left\langle n\right| F\left| m\right\rangle ^{*},$ firstly
we make use of a special
case of Baker-Hausdorff Theorem to rewrite the operator F as a product of
operators i.e we use the equality

$\exp [i\epsilon (a^{+}+a)]=e^{-(\left| \epsilon \right|
^2[a^{+},a])}e^{i\epsilon a^{+}}e^{i\epsilon a}.$

In doing so we have substituted for the commutation
relation of the operators $a$
and $a^{+}$ the value of the commutator given in Eq.(\ref{comm}) in the
limit $q\rightarrow 1.$ This approximation considerably simplifies the
further evaluation of matrix elements of $F$ using the defining Eqs.(\ref
{comm},\ref{vec}) for the operators a, a$^{+}$ and the vectors $\left|
n\right\rangle .$ The final expression for $m\leq n$ is given as,

\begin{equation}
\label{fmat}
\left\langle m\right| F\left| n\right\rangle =\frac{e^{-\left| \epsilon
\right| ^2}(i\epsilon )^{n-m}[m]_q^{\frac 12}!}{[n]_q^{\frac 12}!}%
\sum\limits_{k=0}^m\frac{(\epsilon )^{2k}(-1)^k[n]_q!}{k!(n-m+k)![m-k]_q!}
\end{equation}
Depending on the difference
$n-m$ the matrix element $\left\langle m\left| F\right| n\right\rangle $ can
be real, imaginary, positive or negative.
For the special case of $n=m+1$ and to first order in $\epsilon $ the matrix
element reduces to
$\left\langle m\right| F\left| n\right\rangle =i\epsilon \sqrt{[n]_{q}}.$

\hspace{.5 in}
In the Lamb Dicke regime the model is analogous to
$q$-analog of Jaynes-Cumming Model\cite{Jaynes} with center of mass motion
quanta playing a
role similar to that of the quantized radiation field.
As pointed out in ref.\cite{block}, the effective Rabi frequency
for trapping model can be easily calculated in the lowest approximation.
In a similar way, considering
only those transitions involving single trap quantum exchange we can easily
calculate the analog of the effective Rabi frequency in the deformed
harmonic Oscillator trap. In case the driving field is tuned to the first
vibrational sideband, $\Delta =\pm \omega ,$ all other transitions can be
neglected if these are oscillating at sufficiently high frequencies. For $%
\omega \gg \Omega $, the rotating wave approximation can be used and the
dynamical equations  solved to give the eigenvalues and the
effective Rabi frequency. In ref.\cite{block} the effective Rabi frequency
obtained is $\mu (m)=\sqrt{(\omega +\Delta )^2+\Omega ^2\epsilon ^2(m+1)}.$
For the $q$-deformed oscillator trap the effective
frequency , to first order in $\epsilon $, is a
function of deformation  given by $\mu (m)=\sqrt{[
\frac{\omega}{2}[cosh(2 \tau(m+1))+1] +\Delta]^2
+\Omega ^2\epsilon ^2[m+1]_q}.$

\section{INITIAL CONDITIONS}

\hspace{0.5 in}
The initial state center of mass motion of the atom is represented by $q$%
-analog Glauber Coherent state(GCS), while the atom is in the ground state.
The $q$ -analog of GCS is written as

\begin{equation}
\label{coherent}\left| \alpha \right\rangle _q=\exp {}_q^{-\frac 12\left|
\alpha \right| ^2}\sum\limits_{n=0}^\infty \frac{\alpha ^n}{\sqrt{[n]_q!}}%
\left| n\right\rangle
\end{equation}

with the $q$-exponential defined through

$$
\exp {}_q^x=\sum\limits_{n=0}^\infty \frac{x^n}{[n]_q!}
$$

The complex parameter $\alpha$ determines the average number of trap quanta
associated with the state $\vert \alpha \rangle _q$.

\section{POPULATION INVERSION AND QUASI-PROBABILITIES}

\hspace{.5 in}
In figures (1) and (2), we plot the population inversion as a
function of redefined time parameter
$t(= \frac{\Omega t}{2\pi })$ for $q=e^\tau $ with $\tau $ taking real
values 0.0, 0.002, 0.003, 0.004, 0.006, 0.01 and 0.1 respectively.
Besides that we also use the parameter values $\overline{%
\epsilon }=\frac \epsilon \Omega =50$ and $\overline{\Delta }=\frac \Delta
\Omega =-50$.
For $\tau=0.0$ the results obtained
by Blockley et. al.\cite{block} are reproduced,
when a harmonic oscillator trap is used and mean number of trap quanta
is $\overline{m}=16$.  In the numerical calculation the
maximum value of m in eq.(\ref{psi}) is restricted to $m=32$. In fig.(1)
for $\tau=0.002$,
the first revival is seen to  peak earlier than the one for $\tau=0.0$ and
a well defined collapse appears between the first and the second revival.
For a very small value of $\tau =0.003$ the
collapse and revival pattern is seen to become much more pronounced(well
defined) as compared to the undeformed case. However as the deformation is
increased to $\tau =0.004$ and $0.006$, the pattern again becomes diffuse
after the first revival. The time interval after which the first
revival occurs is seen to become shorter as $\tau$ is increased as
seen in Fig.(2).
Another interesting feature of the collapses and revivals is that
as the deformation increases the collapse period of
the system is seen to exhibit
increasing level of coherent population trapping. For a large deformation,
that in the context of this system is something like $q\geq e^{0.01}$,
the time dependence of popuation inversion is considerably washed away.
For still
larger values of $\tau$ for example $\tau=0.1$ in fig.(2) no time
dependence of the system is seen anymore. Similar results are
obtained when imaginary values of the parameter $\tau $ are used. It is
expected because for very small deformations the values of $[x]_q$ for $%
q=e^\tau $ with $\tau $ real are very close to the values of $[x]_q$ for $%
\tau $ imaginary so long as the modulus of $\tau $ is the same.

\hspace{.5 in}
To obtain further insight into the energy exchange between the atomic
center of mass motion and the internal degrees of freedom of the
atom, we calculate the quasi-probability distributions in the
$\alpha_r-\alpha_i$ plane. The Quasi-probability  or
Q-function is defined
as $Q(\alpha)=(1/\pi)\langle \alpha \vert \rho_{red} \vert \alpha
\rangle_q$, $\rho_{red}$ being the density matrix reduced for
degrees of freedom of center of mass motion. Figure (3) shows  in
the upperv part the
time evolution of Q-function for zero deformation at $t_0=0.0,
t_1=30  , t_2=130$  and $t_3=160$  respectively. Starting
with the center of
mass initial state as a coherent state $\vert \alpha_0 \rangle$,
the initially single peaked
Q-function splits into two peaked function counterrotating in
the complex plane. At $t_3=160 $  , when two peaks collide a revival
of inversion oscillation occurs. A similar plot for $\tau=0.003$
is shown  for $t_0=0.0  ,t_1=30 $ and $ t_2=129.6$
respectively.
An interesting feature of the Q-function for $\tau=0.003$ is the
presence of as many as five peaks at $t_1=30 $. A sequenece of
Q-function distributions,  for $t=5, t=8, t=9, t=10, t=15$ and $t=20$
shows the spreading out and breaking up of the quasi-probability
into as many as eight distinct peaks(for $t=15$.
The revival of Rabi-
oscillations occurs when all  peaks collide together into a single
peak at $t_2=129.6$. We also plot  in Fig.(3) the Q-function
for $\tau=0.004$ at $t=10$
again showing as many as seven peaks.
It is apparent that when trap potential is
q-deformed the quasi-probability distribution is no longer
symmetrically distributed about $\alpha_i=0.0$ line.This asymmetry
 is a manifestation of unequal energy spacings between different
trap states.

\section{CONCLUSIONS}

\hspace{.5 in}
This calculation shows that in a q-deformed oscillator trap for a suitable
deformation the collapse and revival patterns become well defined. As
proposed by Blockley at el.\cite{block} the theoretical predictions in this
case can  be verified experimentally. From our earlier work\cite
{She92,She94} with physical systems we may conjecture that the deformation
corresponds to the presence of some nonlinearitis in the trapping potential.
In this case small scale nonlinearities are expected to play a
beneficial role
facilitating experimental observation. But large nonlinearities can wash out
the collapse and revival pattern completely. This result is very
important
for constructing -an atom in a trap system- with a high probability
for observing collapses and revivals of population  inversion.

\hspace{.5 in}
In the study of interaction of a three level atom with radiation in Ref \cite
{Cardi89} it is shown that by preparing the atom in a special way the time
dependent collapses and revivals are either greatly diminished or vanish all
together. The initial preparation forces the atomic population to remain
coherently trapped in this configuration. Apparently $q$-deformed initial
state is similar to a dressed atomic states. We may conclude therefore that
the nonlinearities of the trapping potential lead to effects similar to
those obtained by setting initial atomic state amplitudes to selected values.

\hspace{.5 in}
In Ref.\cite{Chai90} collapses and revivals have been studied for JCM with
an intensity dependent coupling constant by using $q$-analog of harmonic
oscillator to represent the radiation quanta. In their calculation the
collapses and revivals are seen to become more diffuse as the deformation is
increased. Our result is different from theirs in the sense that the
collapses and revivals become well defined. The major difference in their
calculation and ours is that they use the zero order hamiltonian to be a
harmonic oscillator Hamiltonian whereas we have explicitely included the
zero order spectrum of $q$-analog oscillator.

\acknowledgements

S.S. Sharma would like to thank the Department of
Physics and Astronomy at Rutgers University for its
hospitality and to acknowledge
financial support from CNPq, Brazil.
N. K. Sharma would like to thank the Department of Mathematics
at Rutgers University for its hospitality.

\begin{figure}
\caption{Population inversion versus
$t(={\frac{\Omega t}{2\pi}})$
for $\tau=0.0,\, 0.002,\, 0.003\,$ and $0.004$}
\end{figure}

\begin{figure}
\caption{Same as in Fig.1
for $\tau=0.006, \, 0.008 \, , 0.01, \, 0.1$.}
\end{figure}

\begin{figure}
\caption{Quasi-probability distribution in $\alpha_r - \alpha_i$
plane at $t_0=0.0$, $t_1=30$,
$t_2=130$, $t_3=160$. For $\tau=0.003$, $t_2=129.6$ and for
$\tau=0.004$ all the peaks correspond to $t=10$.}
\end{figure}

\begin{figure}
\caption{
Same as in FIG. 3 for $\tau=0.003$ at times shown in the figure.}
\end{figure}


\begin{references}
\bibitem{Jim}  M. Jimbo, Lett. Math. Phys. {\bf 10} (1985)63; {\bf 11}
(1986)247.
\bibitem{Woron}  S. Woronowicz, publ. RIMS(Kyoto University), {\bf 23}
(1987)117; Commun. Math. Phys. {\bf 111} (1987)613; Invent. Math. {\bf 93}
(1988)35.
\bibitem{Pasq}  V. Pasquier, Nucl. Phys. {\bf B 295}, (1988)491; Commun.
Math. Phys. {\bf 118} (1988)355.
\bibitem{Mac89}  Macfarlane A. J., J. Phys. A{\bf 22} (1989)4581; Wineland
D. J., Itano W. M., Bergquist J. C. and Hulet R. G., Phys. Rev. A, {\bf 36}
(1987) 2220.
\bibitem{Bied89}  Biedenharn L. C., J. Phys. A{\bf 22} (1989)L873.
\bibitem{Died89}  Diedrich F., Bergquist J. C., Itano W. M. and Wineland D.
J., Phys. Rev. Lett., {\bf 62} (1989)403.
\bibitem{Javan81}  Javanainen J. and Stenholm S., Appl. Phys., {\bf 24}
(1981)151.
\bibitem{block}  Blockley C. A., Walls D. F. and Risken H., Europhys. Lett.,
{\bf 17} (1992)509.
\bibitem{Jaynes}  Jaynes E. T. and Cummings F. W., Proc. Inst. Electr. Eng.,
{\bf 51} (1963) 89.
\bibitem{She92}  S. Shelly Sharma, Phys. Rev. C{\bf 46} (1992)904.
\bibitem{She94}  S. Shelly Sharma and N. K. Sharma,
Phys. Rev. C{\bf 50} (1994)2323.
\bibitem{Cardi89}  D. A. Cardimona, M. P. Sharma and M. A. Ortega, J. Phys.
B: At. Mol. Opt. Phys. {\bf 22} (1989)4029.
\bibitem{Chai90}  Chaichian M., Ellinas D. and Kullish P., Phys. Rev.
Letts., {\bf 65} (1990)980.

\end{references}
\end{document}